# Spatial Cognition: A Wave Hypothesis

Robert Worden

Theoretical Neurobiology Group, University College London, London, United Kingdom

rpworden@me.com

Draft 3.7; May 2024

Abstract:

Animals build an internal Bayesian maximum likelihood model of the local 3-D space around them. This 3-D model is essential for controlling all physical movements. There has been large and sustained selection pressure to make it the most precise model possible, given the animal's sense data. A tracking computation has been shown to give a model of space almost as good as the Bayesian best possible model. To do so, it requires precise short-term spatial memory. Neural models of spatial memory have large random errors - too large to support the tracking model.

An alternative to neural spatial memory is described, in which neurons may couple to a wave excitation in the brain, representing the 3-D space. Wave storage of positions can give high precision, fast response, and selective steering of sense data to pattern-recognition modules.

Three lines of evidence support the wave hypothesis: (1) it has much better precision and speed than neural spatial memory, good enough to support object tracking; (2) the central body of the insect brain, whose form is highly conserved across all insect species, is well suited to hold a wave; and (3) the mammalian thalamus, whose round shape is preserved across all mammal species, is well suited to hold a wave. Together, these lines of evidence support the wave hypothesis, compared to the neural synaptic alternative. The wave hypothesis merits further investigation.

# 1. Introduction

Most work in neuroscience assumes that neurons communicate through synaptic connections; that information is represented by firing rates and synaptic strengths; and that only peripheral sensory neurons couple to physical inputs such as light. These are assumed in almost all computational models of the brain since [McCulloch & Pitts 1943; Hebb 1949]; they can be called the **neural synaptic assumption**.

This paper questions that assumption. It proposes that in any animal with complex sense data, some neurons in the brain interact with a wave excitation as transmitters and receivers. The physical nature of the wave is not known. Its computational role is to represent the three-dimensional locations of things around the animal, through a Fourier transform representation of space, as in a hologram. The wave excitation is a master map of the local space around the animal, and is used to plan and control physical movements – which is the first requirement for any animal brain.

Why is there a need for a wave in the brain? Related papers [Worden 2020b, 2024b] describe a working computational model (at Marr's [1982] Level 2) of how animals build a Bayesian maximum likelihood 3-D model of the local space around them – by tracking objects in space as they move relative to the animal. This allows animals to build a 3-D model of space from sense data of lower dimension, such as vision.

The tracking computation works well, allowing an animal to build a precise 3-D spatial model, almost as good as the best possible Bayesian model. Tracking requires a fast and precise short-term memory for the locations of objects. Neural storage of 3-D locations, as stochastic firing rates, has large random errors – especially over small sub-second timescales. The working model of [Worden 2024b] shows that the precision and speed of neural short-term spatial memory is not sufficient to support effective object tracking. The tradeoff between speed and precision cannot be resolved.

Even small animals move rapidly and skillfully, showing that they have good internal models of 3-D space. Something better than neural storage is required to support it. Wave storage of positions can do this – storing the 3-D positions of many independent objects with high precision, low spatial distortion, and fast response; sufficient to hold a precise 3-D model of local space.

Considerations of memory precision and speed, as well as computational efficiency and simplicity, support the wave hypothesis. The paper proposes that wave storage of 3-D locations evolved when animals first had complex sense data and limbs, before the Cambrian era; and that it has been used since then.

There is anatomical evidence for a wave excitation in the brain, in a wide range of species. The paper describes two lines of evidence:

1. **The Insect Central Body** is one of the most complex and least understood parts of the insect brain [Strausfeld 2012; Heinze et al 2021]. Its shape is very well conserved across all insect species. I propose that the shape is conserved because it holds a wave excitation, integrating spatial information from all senses; and that a wave gives the most efficient form of spatial cognition within the limited resources of the insect brain.
2. **The Mammalian Thalamus:** The round shape of the thalamus is conserved across all mammals, and is well suited to hold a wave. In a neural synaptic model of cognition, the anatomy of the thalamus does not make sense. The same synaptic connections could all be made using less brain energy if the thalamus 'exploded' with each thalamic nucleus migrating outwards towards the cortex, reducing the net length and energy costs of white matter. The compact round shape of the thalamus makes sense only if it holds a wave excitation, representing the 3-D positions of things. There is other evidence that a wave in the thalamus holds three-dimensional spatial information.

These three lines of evidence (computational viability, the insect central body, and the thalamus) support the proposal of a wave excitation holding spatial information in all animal brains – strongly enough to make it worth exploring.

The physical nature of the wave is not known. It might be explored in a multi-disciplinary approach, with contributions from biophysics, genetics and proteomics, experimental neuroscience, and other fields.

Researchers may hesitate to search for a wave in the brain, because the idea is unconventional. However, conventional neuroscience has been researched for many years; mainstream research topics are well-trodden paths, long explored by large teams. For a young researcher, looking for a wave in the brain is an unexplored green-fields area, where new ideas can be proposed and where discoveries are to be made. If the wave hypothesis were confirmed, it would be no less than an earthquake in neuroscience. It is worth the attempt.

## 2. The Primacy of Spatial Cognition

Spatial cognition is sometimes seen as one among many different aspects of cognition – alongside others such as memory, learning, pattern recognition, foraging, reproductive behaviour, emotion and language.

Spatial cognition is more than that. It is the first thing that brains need to do; all brains are required to do it very well. It is in some sense the core function of brains, the centre of all other forms of cognition. There are still essentially no working neural computational models of spatial cognition. It is a priority for neuroscience put this right.

Animals are living beings that move at every moment of their waking lives, and they do so very skillfully. The fitness penalties of unskilled movements are always present, and can be terminally serious. Brains need to control every movement, and they can only do this by understanding the locations of things – including the animal's own limbs – in the three-dimensional space around the animal.

The selection pressure on any cognitive function is proportional to how often it is needed in a lifetime, and to the fitness penalties for not doing it well. As spatial cognition is essential at every moment of the day, there has been very large and and sustained selection pressure on animal brains since the Cambrian era, to do better 3-D spatial cognition. The evidence is that they do it very well. Spatial cognition is the core requirement that has shaped animal brains; it is the primary function of a brain. It is an urgent requirement to understand how spatial cognition works.

## 3. Requirements for Spatial Cognition

There is a common set of requirements for spatial cognition in any animal which has complex sense data – for any animal which has powerful eyes, or capable limbs. These requirements are:

1. **To represent positions in three dimensions:** The space around any animal is three-dimensional, and movement is movement in three dimensions. It follows that a faithful internal model of reality needs to be three-dimensional.
2. **To represent positions precisely:** A more precise internal model of reality leads to better choices of action and greater fitness. The resolution of any species' eyes gives one measure of the precision required in its brain; there would be little purpose in having an eye with high spatial resolution, if the brain could not represent 3-D reality with the same precision.
3. **To represent positions over a wide range of distance scales:** Any species needs to represent reality over a very wide range of distance scales – from the large distances involved in locomotion and predator avoidance, down to the smallest distances involved in fine control of limbs, for manipulation and recognition of small objects.
4. **Rapid response to change:** Changes in local space can happen in a fraction of a second. An immediate and appropriate response may be needed, in order to survive.
5. **Multi-sensory integration:** Information about any external thing comes from several senses including vision, proprioception and sound. The brain needs to integrate these different sources into one maximum likelihood model of the world.
6. **Detection of Motion:** It is important for animals to detect what is moving in their surroundings; for instance, it may be food, or may be a threat. To do this when moving, direct use of the visual field is of limited utility, because of apparent motion in the visual field caused by the animal's own motion. A 3-D model in allocentric space is a gold standard for determining true motion of things.
7. **Spatial Steering and Binding:** To classify things in space as accurately as possible requires the integration of all information coming from the same spatial location, because it is likely to come from the same thing. Information of different sense modalities needs to be bound together; it needs to be routed or steered to some recognition module in the brain.
8. **Classifying objects and events:** To decide what to do next, an animal needs to rapidly classify things or events around it into types, using sense data of all modalities – to know what it can do with any thing, or what the thing might do to it.
9. **Fast learning:** The classifications of things and events are not built into the brain by natural selection; they must be learned within the short timeframe of the animal's life. This requires learning from a small number of learning examples - unlike the slow learning done by deep neural nets.
10. **Spatially invariant learning:** An animal must learn to classify any object, wherever that object appears in its surroundings. It must learn in a spatially invariant manner - from examples which occur in different places and orientations relative to itself.
11. **Fast learning of effective movements:** Motion is too variable to be fully innate. Movements must be learned rapidly, within a small part of the animal's expected lifetime.

These requirements exert strong selection pressures on the evolution of brains. It is not sufficient for brains just to do them to some basic standard; brains must do them as well as possible. They are competitive requirements; it is not sufficient just to meet a requirement 'well enough'. If species A meets a requirement better than species B, then A will out-compete B, and B will not survive. Evolution is an arms race for brains. We expect these requirements to be very well met; and empirically, they are.

The requirements (1) – (5) relate to understanding 3-D space to control physical movements – something animals do very well, because a failure to move skillfully could cause death or injury at any moment, so there is very strong selection pressure to do it well. The requirements (6) – (11) relate to a broader understanding of their surroundings – which may not lead to such very high selection pressures, but nevertheless gives rise to strong competitive selection pressures.

Even insect brains, which typically contain less than a million neurons, meet the requirements (1) – (11), and meet them well. We are not just looking for brain designs that meet the requirements; we are looking for designs which meet them efficiently and meet them well, using small numbers of neurons.

## 4. An Object Tracking Model of 3-D Spatial Cognition

The Bayesian cognition hypothesis, that animals build Bayesian internal models of the world from their sense data, [Knill & Pouget 2004; Friston 2003, 2010] has been successful in many cognitive domains. Bayesian cognition gives the best possible fitness, as can be shown from evolutionary arguments. In [Worden 1995, 2024a] I showed that the computational requirement for any animal brain (at Marr's [1982] computational level 1) is to choose actions as if using an equation, the Requirement Equation. This equation resembles Bayes' theorem, with extra factors depending on actions and their fitness payoffs. The resulting choice of actions gives greater fitness than any other choice, and so the evolution of brains will converge towards this form of cognition. In complex domains, it requires animals to build Bayesian maximum likelihood models of the world.

For spatial cognition, it requires animals to build a 3-D spatial model of the space surrounding them, using all available sense data, and obeying the constraints of Euclidean geometry, kinematics and physics. These constraints have been true for all evolutionary time, and have exerted constant selection pressures on brains throughout all that time; so the constraints can be expected to be reflected in animal brains with high precision. Animal movement exploits the laws of physics very well; it must be controlled from a model of space which obeys those laws.

Full Bayesian computation of a maximum likelihood spatial model is not tractable in animal brains in real time, but it can be computed on digital computers. [Worden 2024b] describes a working computation of a Bayesian optimal 3-D spatial model, for bees (using vision and Structure from Motion) and bats (using echo-location and Structure From Motion). This model, at Marr's [1982] level 2, requires the animal to use not just current sense data, but to combine it with recent sense data to build the 3-D model of space. This allows the construction of an accurate spatial model, with precision comparable to (and sometimes better than) the animal's sense data. It is expensive to compute, and it is not likely that animals compute full Bayesian 3-D spatial models.

[Worden 2024b] also describes a dynamical object tracking model for bees and bats. In this model, the animal does not retain recent sense data, but only uses current sense data (vision or echo-location) to update its most recent 'tracking' estimates of object positions, in an allocentric frame of reference, where most objects are stationary. This is easier to compute than the full Bayesian 3-D model, and gives a model of very similar precision and quality. It is an example of computing shape from motion (SFM) [Murray et al 2003]. It is a possible candidate for how animals model the space around them, to control their movements.

While the tracking model is fast and economical, it requires the storage of recent tracked position estimates; so it requires a three-dimensional spatial short-term memory. The model program shows that this short-term memory needs to have high spatial precision (errors in spatial displacements of the order of 1% or less) to support effective tracking, and to support functions which depend on tracking, such as the detection of moving objects.

There has been little consideration of how a neural short-term 3-D spatial memory could do this. As discussed in the next section, it is likely that neural spatial memory would not have sufficient precision to support the tracking model.

## 5. Neural Spatial Memory: Speed and Precision

How could neural firing rates represent three-dimensional information, as is required for short-term neural spatial memory? I consider three options:

A. Represent two of the dimensions by position in a 2-D neural sheet, and represent the third dimension by firing rates at locations in the sheet – like a visual cortex with depth.
B. Represent all three dimensions of any position by neural firing rates.
C. Represent all three dimensions by positions of active neurons in a 3-D 'clump' of neurons.

Option (C) is unattractive for several reasons. To give high spatial resolution, it would require to be a large and prominent clump of neurons, such as has not been observed in brains. It would also pose serious problems of neural connectivity, for neurons in the middle of the 3-D clump.

Option (A) potentially gives high resolution in two of the dimensions, represented by positions in the sheet. A possible drawback is that for motion detection, it would be useful to represent space in an allocentric frame of reference, where most things do not move; but the visual cortex does not use such a frame, so the spatial memory would need to be somewhere else in the brain.

Option B raises questions about how the three dimensions are defined (in what kind of coordinate system?); and then, how commonly used operations, such as vector addition of displacements, would be computed. It is hard to devise a simple representation of three-dimensional space by neural firing rates, and then to use it to do spatial computations simply. This issue of computational complexity also arises in option (A); there is little to say about it, except that complex forms of processing appear to be needed, and that they have not yet been considered.

In the tracking model, memory noise as little as 1 part in 40 degrades motion detection down to near-random levels, and the precision required to support tracking is of the order of 1 part in 100. Could a neural spatial memory give these levels of precision?

Quantities such as components of vectors can be represented by stochastic neural firing rates. If a single neuron fires stochastically with N action potentials per second, in one second it represents information with precision approximately one part in $\sqrt{N}$. To get a precision of the order of 1% (as appears to be required to support motion detection) in one second would require $N = 10,000$, which is an unrealistically high firing rate. Typical neural firing rates are 5 - 50 pulses per second.

The problem is more serious because animal brains need to choose physical actions faster than once per second. For a small mammal, the required times may be of the order of 100 milliseconds. For insects, whose vision is 5 times faster than our own [Chittka 2022], the timescales may be a few tens of milliseconds. It is only possible to fit a small number of neural action potentials into this time – giving very poor precision.

To represent spatial coordinates by stochastic single neuron firing rates requires an impossible tradeoff between speed and precision. It cannot be done.

We must therefore look for other solutions. One possible approach is parallelism, to get higher firing rates. Using the numbers above, a very high degree of parallelism would be required to get the required aggregate firing rate – perhaps 100 parallel neurons to represent one dimension of one position. As we know that brains represent the positions of many objects at once, this soon scales up to prohibitive numbers of neurons – particularly in insect brains, which have fewer than a million neurons in total.

Another approach would be to use some non-stochastic firing pattern, such as regular bursts. Another approach would be to use a more complex encoding of distances, perhaps using small linked groups of neurons to encode position coordinates in multiple firing rates.

A drawback of all these approaches is that they make an already complex problem – how to do spatial computations, such as vector additions – yet more complex. Some of the possible approaches would cause distinctive neural connectivity or firing patterns, which would have been observed.

It appears therefore that neuronal short-term memory cannot give sufficient precision and speed to support three-dimensional object tracking (or any similar model of 3-D object location), at the levels of precision and speed which small animals routinely attain. It is of course possible that someone will propose better forms of neural spatial memory, to overcome these problems.

## 6. Wave Storage of Spatial Information

This section describes an alternative to a neural implementation of spatial memory.

For any computation, if the physical computing mechanism matches the physics of what is computed, the computation can be simpler, more economical and more precise. You can 'let the physics do the computing' – directly, rather than building some complex device to compute indirectly. This principle was responsible for the early uses of analogue computers, before digital electronics became predominant.

The same principle can be applied in spatial cognition. If there is some approximately spherical volume in the brain, which can hold wave excitations, then each wave can have a different wave vector, or **k**-vector. This is a three-dimensional vector which describes both a wavelength and a direction of wave motion (orthogonal to the wave front). If the physics of the wave is linear, the same volume can hold many independent waves, with different **k** vectors – which do not interfere with one another.

A wave can be used to store the independent locations of many objects – objects with different locations **r**, related to the wave vectors by **k**=α**r**, where α is a constant. If we assume that:

- Each wave excitation can persist for short periods (fractions of a second)
- The minimum possible wavelength λ is small compared to the diameter D of the volume (so there is a large range of possible **k**-vectors, in all directions)
- Neurons can couple selectively to waves of different wavelength and direction, as transmitters and receivers (e.g. one neuron might have its wave receptors or transmitters aligned and spaced with the wave fronts, to be selective near one **k**-vector) then neurons could use the wave as a short-term memory for the locations of many objects. This form of spatial memory could have major benefits over storage in neural firing rates:

1. The three dimensions of the wave vector correspond directly to the three dimensions of object positions; there is no need for any preferred direction, or for the representation to be asymmetric between directions. There is no need to choose a coordinate system. It is a simple and direct representation of positions.
2. A large number of **k**-vectors (independent object positions) - of the order of $(D/\lambda)^3$ - can be stored in the same wave volume.
3. The stored precision of each object location in any dimension is approximately one part in $(D/\lambda)$ – which can be better than one part in 100, as appears to be required for effective motion detection.
4. As in a hologram (which works by the same principle) there is very little spatial distortion of positions.
5. In principle, the wave can be updated very fast, say within a few milliseconds. There is no hard tradeoff between speed and precision.

These are major benefits, possibly overcoming the serious problems with neural storage described in the previous section. They are enough to make the tracking model of spatial cognition workable – which it seems not to be, with purely neural storage of positions.

If spatial positions are stored in a wave in the brain, there must be some minimum possible wavelength $\lambda_{min}$ that neurons can couple to; which implies that there is a maximum **k**-vector, and a maximum distance that can be represented. This is a problem for representing very large distances, which animals sometimes need to do. The wave excitation cannot represent Euclidean space directly, but must represent some transform of Euclidean space, designed to minimise geometric distortions. In this respect, projective transforms of space [Rudrauf et al 2017, 2022] are particularly useful, as they preserve straight lines; and straight lines are important for controlling motion and recognizing shapes. So the wave storage may use some near-projective transform of Euclidean space.

This may be why we see the stars as a spherical canopy, and why perception has minor distortions of Euclidean geometry, especially at large distances.

Storage in a wave has potential for a spatial short-term memory, which could be greatly superior to neural memory. Many details remain to be worked out – including:

a) What is the physical nature of the wave?
b) How can the wave persist for the necessary times?
c) What is the source of energy for the wave?
d) How do neurons couple to the wave? What genes and proteins are involved?
e) Can neurons have steerable coupling to the wave?
f) Where in the brain does the wave reside?

On question (a), while the wave might be electromagnetic [McFadden 2002] there are other possibilities, which could be better insulated from neural background noise. There are possible answers for question (f), described in the following two sections.

## 7. The Insect Central Body

Insect brains typically have volume less than one cubic millimetre, with fewer than a million neurons – yet insects have a range of complex and adaptive behaviour. It is no longer believed that insects are only capable of stereotyped pre-programmed actions, or that they cannot make fine discriminations about their surroundings. They can do these things, and do it very fast [Chittka 2022].

It follows that insects have capable spatial cognition, meeting the requirements of section 3 within a very small brain.

The insect central body is, as its name implies, a small central region found in all insect brains. It has been much studied and is linked with a wide variety of functions, including navigation and locomotion. It is innervated by neurons originating from nearly all sensory modalities, and it is thought to play a role in multi-sensory integration. Because of its important roles and its complexity, Strausfeld [2012] has called it a 'brain within a brain'. While its neural circuitry has been extensively studied, there are few computational models of its functioning.

The central body has features which may imply that it is the site of a wave excitation in the brain. If it is so, that is consistent with a role for the central body as a spatial blackboard of the insect brain, routing and steering multi-sensory information between other parts of the brain, to assist in building a Bayesian maximum likelihood 3D model of the insect's surroundings.

The main evidence for a wave is the remarkable constancy of the shape of the central body across all insect species, some of which diverged from others over 400 million years ago. This constant shape would not be expected if the role of the central body was just to hold synaptic connections, because many different and distorted shapes could give the same neural connections. One would expect the central body, like most parts of insect brains, to have different shapes in different species, to optimize the use of space and to minimize neural connection lengths according to species-specific requirements and neighboring brain regions.

A shape which is highly preserved across species may imply that the shape serves some physical purpose, other than the connection of synapses. One purpose is to act as a container for a wave excitation in three dimensions. If the excitation needs to hold comparable numbers of waves in all three dimensions (to represent all three dimensions of positions with comparably good spatial resolution), one would expect the shape to be approximately spherical – as the insect central body is.

The central body (CB) consists of the Fan-shaped body (FB) which is next to, and partially surrounds, the Ellipsoidal Body (EB). These can be seen in a range of histological preparations. All preparations show a remarkably constant central body shape across insect species, as is illustrated in [Strausfeld 2012]. Figure 1 from [Strausfeld 2012] shows the central body of a honey bee brain.

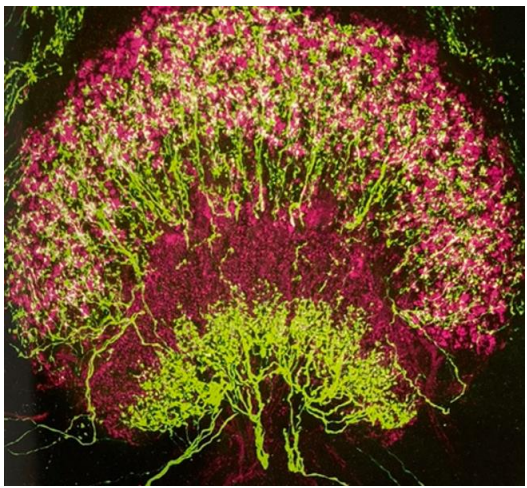

*Figure 1: Central Body in the brain of a honey bee*

The central body can also be viewed in three dimensions for approximately 20 insect species in the insect Brain Database [Heinze et al., https://insectbraindb.org]. 3-D models of insect brains and brain parts can be downloaded from this site.

The outline shapes of the FB and EB, from the database, are shown side by side for 8 species, in figures 2 and 3:

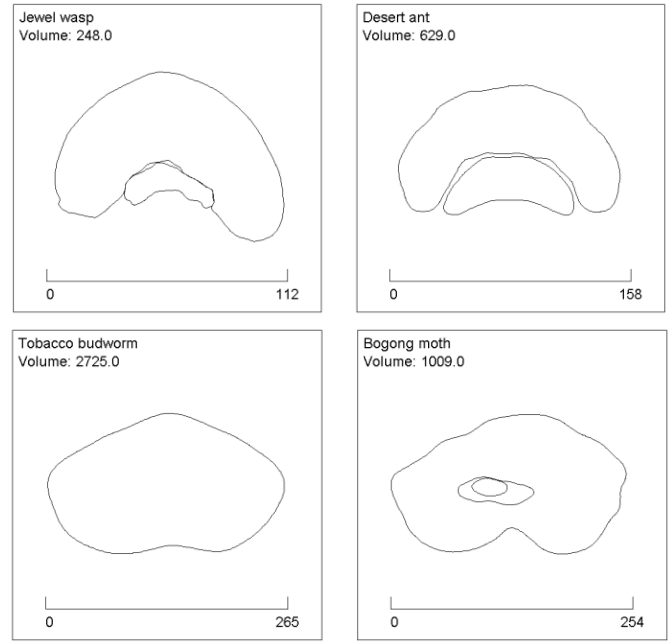

*Figure 2: Outlines of the FB and EB in four insect species, from the Insect Brain Database [Heinze et al 2023]. Units of length are microns ($10^{-6}$ M), and volumes are in units of 1,000 cubic microns*

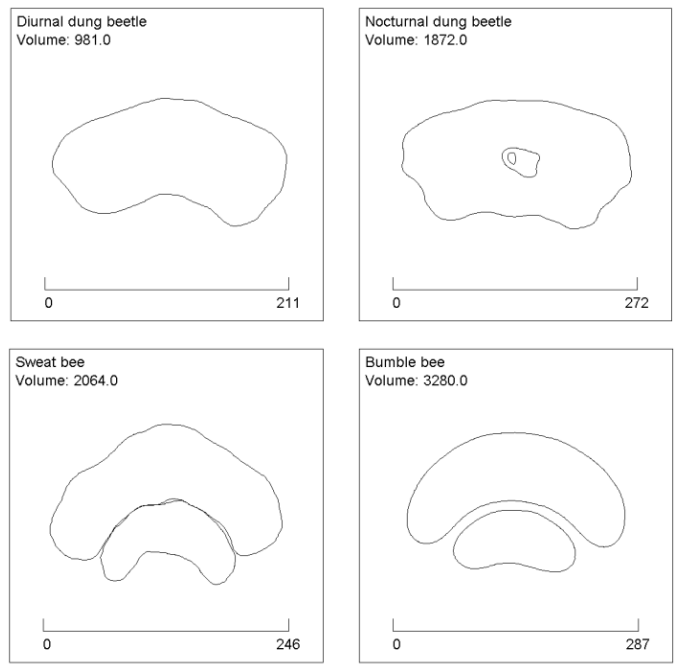

*Figure 3: Outlines of the FB and EB in four insect species, from the Insect Brain Database [Heinze et al 2023] . Units of length are microns($10^{-6}$ M), and volumes are in units of 1,000 cubic microns*

Species are shown in order of increasing total brain volume, from top to bottom. Total brain volume goes from 17,000,000 cubic microns (Jewel Wasp) to 660,000,000 cubic microns, or 0.7 cubic millimetre (Bumble bee).

For each species, the volumes of the central body are shown, in units of 1,000 cubic microns. Typically, the central body only occupies between 0.5% and 2% of the total insect brain volume; other parts of the brain, such as the optic lobes and the mushroom bodies, are much larger.

The pictures show a two-dimensional slice of each CB. To quantify the three-dimensional shape constancy across species of the central body, I have defined a measure of the shape constancy of any brain part between any two insect species, by the procedure:

- Normalise the brain part, scaling it so it has the same volume in both species.
- Rotate the part for each species so that they both have the same orientation, defined by the major and minor axes of their moment of inertia tensors (the second moments of the volume distribution). This orients them as shown in the diagrams, with major axis from left to right (x), the next major axis vertically (y), and the smallest axis (z) towards the viewer.
- Measure the shared volume S which is inside the brain parts for both species, and the individual volume I which is in a part for one species but not in the other.
- There is a four-fold ambiguity, of a 180° rotation about any axis. Choose the ambiguity to minimise I.
- The measure of shape similarity is I/S.

Using 3D models from the Insect Brain Database, I have measured the shape constancy of the central body for 18 species, taking the average of I/S for all pairs of species. For comparison, I have made the same measure for some other parts of the insect brain. The results are shown in the table:

| Brain region | Number of species | Mean disparity I/S per species pair (percent) |
|---|---|---|
| Fan-shaped body + Ellipsoidal body (central body) | 15 | 26 |
| Parabrachial Bridge (central complex) | 14 | 257 |
| Peduncle (mushroom body) | 15 | 184 |
| Medial lobe (mushroom body) | 13 | 129 |
| Calyces (mushroom body) | 11 | 95 |
| Lamina (optic lobe) | 8 | 145 |
| Lobula (optic lobe) | 11 | 44 |
| Lobula plate (optic lobe) | 10 | 55 |
| Medulla (optic lobe) | 14 | 27 |
| Antennal lobe | 16 | 40 |
| Lateral accessory lobe | 8 | 42 |
| BU (lateral complex) | 10 | 41 |

*Table 1: Shape disparities for different parts of the insect brain, computed from the models in the Insect brain database. The number of species varies with the brain part because not all species have data for all brain parts.*

The table confirms numerically what is evident visually – that the central body, while not having an entirely constant shape across species, is more constant in shape than any other part of the insect brain – with one possible exception.

The only other part of the brain which has similar constancy to the central body is the medulla in the optic lobe – which, interestingly, has a similar shape to the central body. Being closely related to the eye, it probably has a mainly retinotopic neural organization.

Neurons in insect brains are often unipolar, having a cell body with one emerging process branching to axon and dendrite. Many of the cell bodies are external to the bulk of the brain. There are very few cell bodies in the central body; but it is innervated by many different types of neuron, implying that it has complex and important functions.

Many of the neurons that innervate the CB have synapses widely distributed through its volume. If these synapses are closely associated with some sub-neural unit which transmits or receives a wave, then the widely distributed units can define a wave vector with high spatial resolution. This is a requirement for representing three-dimensional space with high resolution. Widely spaced synapses are consistent with representation of space by a wave.

Note that the linear dimensions of the central body vary only by about a factor 2.5 (from 112μm to 287 μm), while total brain volumes vary by a factor of 40. In the wave model, it is the linear dimension of the central body that determines its spatial resolution. The fairly constant sizes of the insect central body are consistent with the wave model.

## 8. The Mammalian Thalamus

The thalamus occupies a central position in the mammalian brain, and is richly interconnected to many cortical areas. It has been proposed that the thalamus plays a role in integrating the functions of those areas, acting as a blackboard [Erman et al 1980; Nii 1986; Baars 1988; Llinas & Anthony 1993; Mumford 1991; Lee & Mumford 2003; Worden, Bennett and Neascu 2021].

This section explores the hypothesis that the thalamus holds a wave excitation, as was proposed for the insect central body in the previous section. Several pieces of evidence support this hypothesis. Some features of the thalamus cannot be accounted for without a wave.

The first evidence is the shape of the thalamus. Across many species, the thalamus has a well preserved and near-spherical shape [Sherman & Guillery 2006; Jones 2007] – a form which is well suited to holding a wave excitation, with comparable numbers of wavelengths in each of its three dimensions. This regular near-spherical shape contrasts with other parts of the mammalian brain, such as the hippocampus or the cerebral cortex. Their irregular shapes are sufficient to hold neural connections, which depend only on connection topology, not geometry. Most parts of the brain have irregular and species-specific shapes, dictated by space requirements and their impinging neighbours. The thalamus is a striking exception, and it must be for a reason. That reason may be the need to hold waves.

The shape and volume of the thalamus in several species is shown in figure 4, taken from [Hailey & Krubitzer 2019].

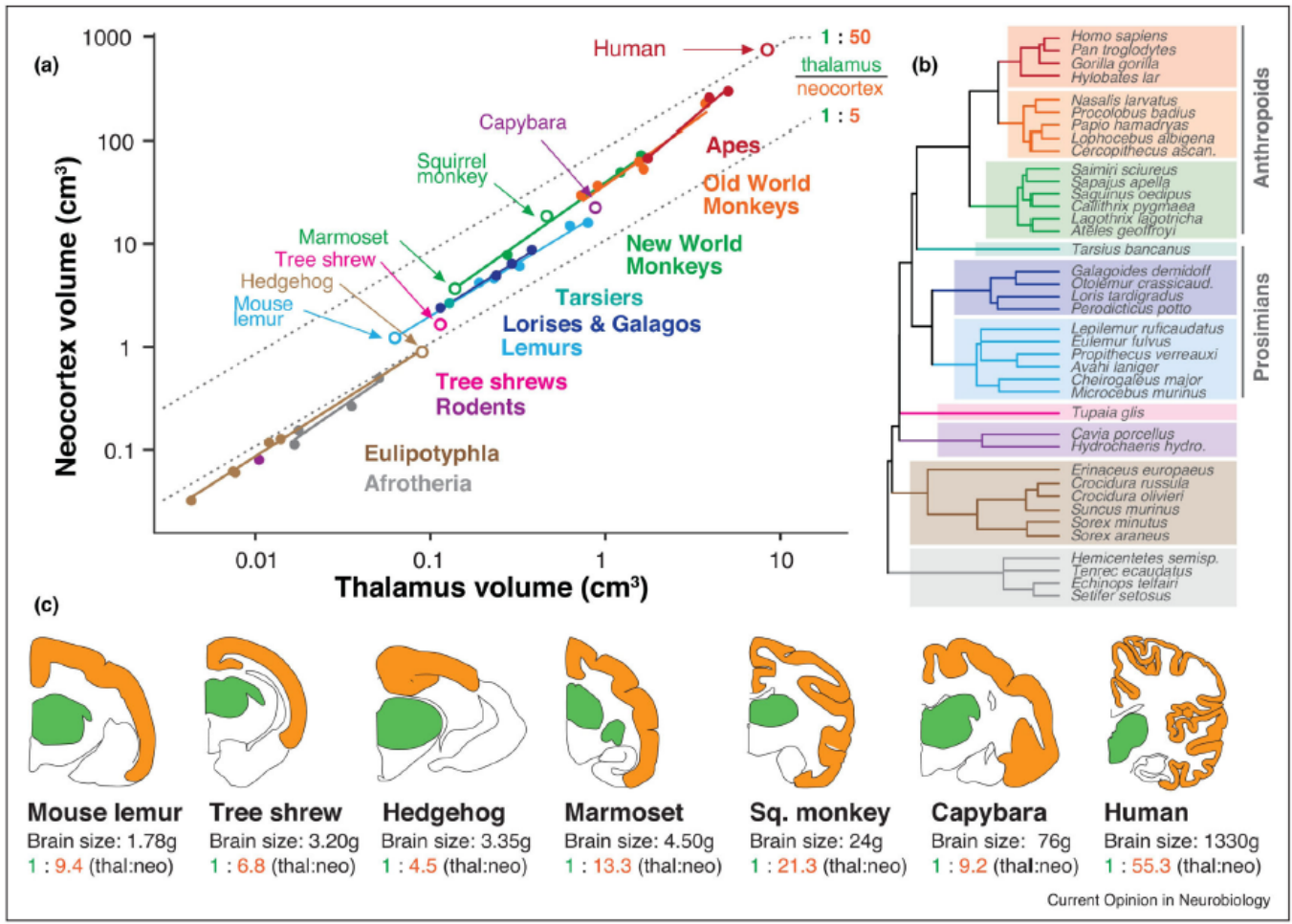

*Figure 4: Shape and size of the thalamus in various species, from [Hailey & Krubitzer 2019]*

With the apparent exception of the marmoset, the visible shapes of the thalamus (green) are consistent with each thalamus being approximately spherical, with the two thalami joined at the centre line of the brain. This is in sharp contrast to the convoluted variable form of the neocortex (orange)

The second piece of evidence comes from considerations of brain energy consumption, through the lengths of axons.

The thalamus has rich two-way connections to almost all parts of the cortex [Sherman 2006; Halassa & Sherman 2019], and those connections have an energy cost – proportional to the aggregate length of the myelinated axons. The thalamus itself consists of a number of thalamic nuclei – and the neural interconnections between thalamic nuclei are weak or non-existent [Sherman & Guillery 2007]. Why do thalamic nuclei need to be close to each other, if they have no interconnections? The aggregate lengths of axons between the thalamus and the cortex could be reduced (and brain energy saved) if the different thalamic nuclei were to separate, so each thalamic nucleus could migrate outwards towards the cortex – reducing aggregate axon lengths. All neural synaptic connections would be preserved – so in a neural synaptic model of cognition, the same computations could all be done.

Figure 4 shows that in all species except possibly the hedgehog, there is a gap between the thalamus and the cortex – a gap which is traversed by cortico-thalamic and thalamo-cortical axons. There are potential energy savings from reduced axon length, if thalamic nuclei were to migrate away from one another, outwards towards cortex. This is shown in figure 5 below.

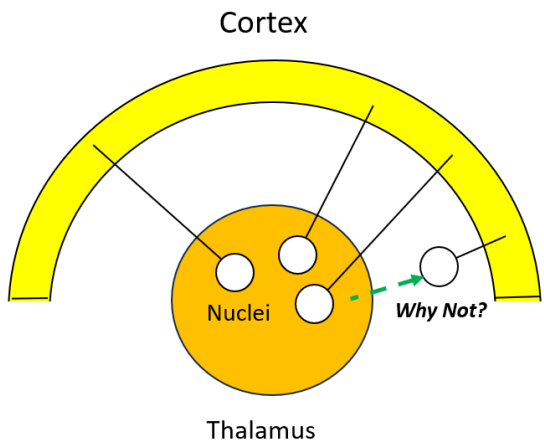

*Figure 5: As individual thalamic nuclei have no connections to each other, each nucleus could migrate outwards towards the cortex, reducing net axon length and still having the same neural connections. Nuclei need to stay together to be immersed in the same wave.*

In the case of the human thalamus, [Worden 2010] has calculated the possible energy reduction from this 'exploding thalamus' and it is a significant energy saving. Brain energy costs for mammals are very important; so if they could be reduced by exploding the thalamus, evolution would surely have done so. Why do the thalamic nuclei stay together?

There is one good reason for thalamic nuclei to stay together – because they all need to be immersed in the same wave excitation. The central position and form of the thalamus is strong evidence for the wave hypothesis.

On a neural synaptic model of cognition, the division of the thalamus into nuclei which hardly connect to one another is ill-designed to assist in any integrative function of the thalamus, like a blackboard – whereas in the wave model, different thalamic nuclei communicate through the wave.

A third line of evidence in support of the wave hypothesis comes from the thalamic reticular nucleus (TRN). This has a very distinctive shape, of a thin shell surrounding the dorsal thalamus. In terms of brain energy consumption, this shape does not make sense. Energy could be saved if, instead of being an extended thin shell, the TRN was more compact and moved some distance towards the centre of the thalamus [Worden 2014].

However, in the wave hypothesis, the shape of the TRN may make sense. A thin shell, surrounding the volume containing the wave, may be a transmitter or receiver of the wave. It is hard to find any other good reason for the shell-like form of the TRN.

Evidence that the role of the thalamus is largely concerned with spatial cognition also comes from the kinds of sense data passing through the thalamus. The thalamus is a relay for sense data of nearly all modalities on its way to the cortex – that is, for all sensory modalities except olfaction. Olfactory data is of little use in rapidly finding the precise locations of things in three dimensions – which may be why it does not pass through the thalamus. This links to the wave hypothesis because, as described in section 7, a wave excitation is well suited to represent three-dimensional spatial information.

Figure 4 shows that for most mammalian species, the thalamus occupies only a small percent of total brain volume, between 1/5 and 1/50 of cortical volume; and the volume of the thalamus scales as (cortex volume)$^{0.8}$ [Hailey & Krubitzer 2019]. This is consistent with the thalamus carrying out its function efficiently - with the required resources scaling only modestly with increasing requirements, as they do in the wave model, but not in the synaptic model. As with the insect central body, the linear dimensions of different species thalami are broadly consistent with the wave hypothesis.

Further evidence comes from the well-documented relay function of the thalamus. Why is the thalamus a relay for most kinds of sense data? What value is added by its relay function? How does it repay the extra energy costs of sense data having to take a longer path to the cortex through the thalamus, and the extra delay of relay neurons firing?

The wave model gives a good rationale for the thalamic relay function. If the thalamic wave holds a 3D map of local reality, it is important to update the map as fast as possible. When sense data arrives at a primary relay of the thalamus – such as the Lateral Geniculate Nucleus (LGN) – it is not only relayed onwards to the visual cortex, but it is also used to update the wave model immediately, using a pre-existing estimate of depth for that region of the visual field. This means that the wave model of local 3D space is updated by incoming sense data as fast as it possibly can be. That is the reason why incoming sense data is diverted through the thalamus on its way to the cortex. This is illustrated below in figure 6.

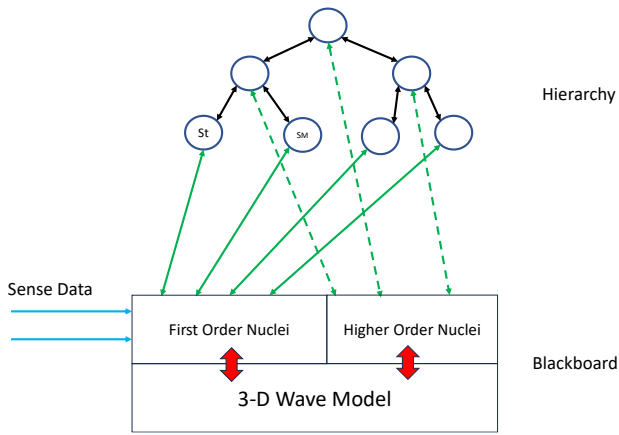

*Figure 6: Roles of thalamic nuclei in spatial cognition. Green arrows represent cortico-thalamic and thalamo-cortical axons. Red arrows denote transmission and reception of the thalamic wave, representing 3D space.*

In this figure, incoming sense data arrives at first order thalamic nuclei such as the LGN. From there it is passed to knowledge sources at the lower levels of the cortical hierarchy, such as stereopsis (St) and shape from motion (SFM). These knowledge sources estimate the depth of the stimulus, which is passed back (solid green arrows) to transmitters in the LGN, which create the wave representation of space (left-hand thick red arrow) from incoming visual data. Higher order thalamic nuclei such as the Pulvinar have similar two-way links to higher levels of the cortical hierarchy (dashed green arrows), and contribute to the wave model (right-hand thick red arrow).

There is an approximate anatomical consistency between primary thalamic nuclei, such as the LGN, and the wave model of section 9. The LGN has laminae of neurons with retina-like organization in each lamina. If each neuron in a lamina controls a transmitter region, the orientation of overlapping transmitter regions in the lamina is approximately that needed to project a wave to the centre of the thalamus, and control its two dimensions by its source in the retina, with the third dimension (depth) controlled by frequency tuning of the transmitter. Some phase steering of transmitters is still needed. The presence of two or more laminae (as in most mammals) may help to project the wave towards the centre of the thalamus, rather than away from it. – like a phased array antenna

The ultrastructure of the thalamus is notable for the presence of synaptic glomeruli – clusters of synapses, which are consistent with multiple synapses playing a role in steering wave receptor units in the thalamus by variable phase delays.

The complexity of thalamic glomeruli suggests that some complex computation is done inside them, and they are not just passive relays between neuron cell bodies which do the computational work, as in the McCulloch-Pitts-like model of the neuron [McCulloch & Pitts 1943]. Figure 7 shows a 3D reconstruction of a glomerulus from electron microscopy, from [Spacek & Lieberman, 1974]

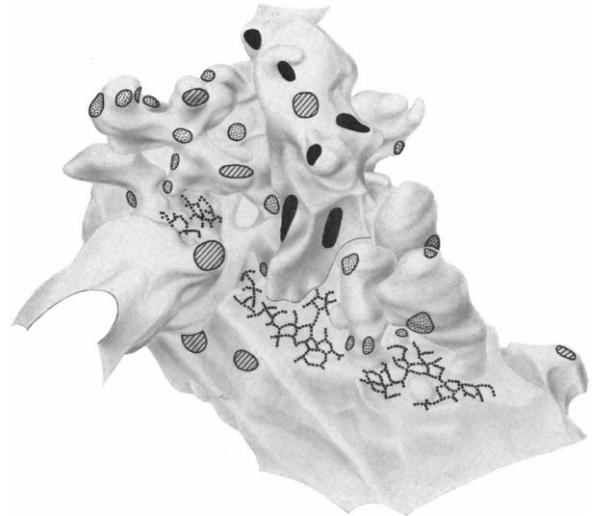

*Figure 7: Graphic reconstruction of a thalamic glomerulus, from [Spacek & Lieberman 1974]*

## 9. Spatial Steering and Binding

This section focuses on the requirements (7) – (10) of section 3:

7. Spatial Steering and Binding
8. Classifying objects and events:
9. Fast learning:
10. Spatially invariant learning:

Taken together, these requirements are very important, and they depend on an internal model of 3-D space. Animals need to know not just the locations of things in space, so they can move skillfully in relation to them. They also need to know what things are – to infer what they are likely to do, or can be used for. Animals need to classify the objects around them into types. This is a pattern recognition problem, and it is spatially invariant pattern recognition. The same object must be recognized, wherever it is in the local space around the animal.

Spatially invariant pattern recognition could be done by a brute force neural net approach, but it would be very inefficient [Denker et al 1987] – particularly in the very large learning times, typically requiring many more learning examples than an animal experiences in its lifetime. Animals need something much more efficient and rapidly learnable. It appears (e.g. from cortical neuroanatomy) that they use small pattern learning and recognition modules, located in different parts of the brain.

This leads to the requirement (7), for spatial steering and binding [Treisman & Gelade 1980; Treisman 1998; Feldman 2013]. How are all the sense data, needed to recognize some object, steered from sense organs to a small learning module? How are sense data of different modalities routed and bound together, whatever their origin in space?

The sense data that need to be routed and bound together all come from the same region in 3-D space, because an object is likely to occupy some small region of space. So spatial steering is required. It is likely that the animal's internal 3-D spatial model plays an important role in sensory signal steering and binding. How does it do so?

There are purely neural models of signal steering [Olshausen et al. 1993, 1995] , by branching switchable connections between neurons. This is illustrated (for one spatial dimension) in figure 8 below.

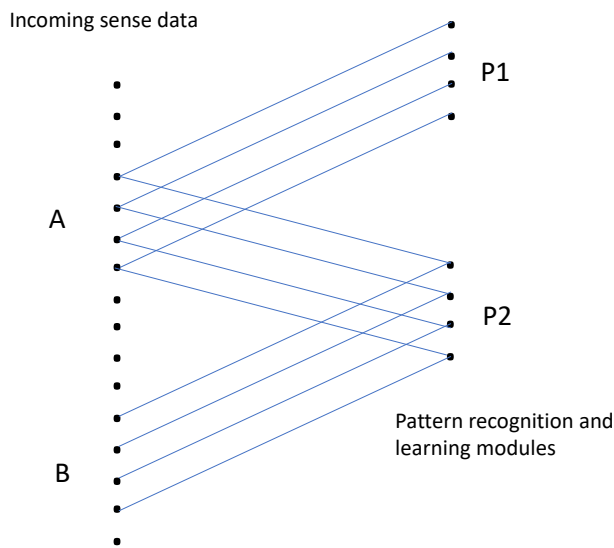

*Figure 8: A Fan-in/fan-out design for spatial steering and binding.*

The line of points on the left denotes sensory inputs – say, a one-dimensional visual field. A and B denote regions in the field. The groups of points P1 and P2 on the left each denote a pattern recognition and learning module. The lines across the diagram denote selective pathways of sense data to modules. Somewhere along these pathways there are switchable neurons or synapses – which are switched off for nearly all the time, but can be switched on to make a selective path from sense data to a pattern recognition unit. (having a neural pathway which is switched off for nearly all the time is in itself inefficient)

This is a fan-in/fan-out design because:

- Sense data needs to fan out from region A to any module such as P1 or P2 which may recognize a pattern in the data.
- Sense data needs to fan in to any module such as P2, from different regions such as A or B in the input field.

While these architectures have been proposed, in one and sometimes two dimensions, the example models have generally been on a small scale, and have not addressed the issues of scaling (in the numbers of neurons and switchable synapses) which would be required to apply the models to the whole 3-D spatial model. I suggest that these issues of scaling are very serious, and rapidly lead to prohibitive numbers of neurons and synapses – especially in small animals such as insects. Yet animals do spatially-invariant learning and pattern recognition very well. For instance, bees rapidly learn and use patterns of flowers in their local habitat.

Wave storage of spatial information offers an alternative to neural steering [Worden et al 2021]. It is fundamental to the wave model that both for input to the wave (of sense data) and output of the wave (e.g. to pattern recognition modules) neurons need to couple to the wave selectively, being sensitive only to a small range of wave vectors. Each neuron may be like an antenna immersed in the wave. It is selective for some wave vector if its dendrites contain small transmitter/receiver units (possibly, synapses) spatially distributed to match the wave vector – to add together the peaks of the wave, possibly in a phase-sensitive manner.

Since neurons must act as selective antennae in the wave, it is possible for them to act like a steerable selective antenna. This could happen if the phase of the contributions of transmitter and receiver units was controlled by a separate steering signal. For instance, the compound synapses or glomeruli in the thalamus could receive a steering signal, which alters their receptive properties.

For spatial steering and binding, a neural synaptic architecture appears to be complex and inefficient, and to have severe difficulties of scaling. The wave architecture requires novel capabilities to make steerable neural antennae, but is simpler and avoids the severe scaling problems.

In an analogy, the superior scaling of the wave model is like the benefit of radio over wired telegraph. Far fewer connections are needed.

## 10. Collecting the Evidence for Wave-based Spatial Memory

This section brings together the main pieces of evidence related to the wave hypothesis of spatial memory.

1. Animals move skillfully and recognize objects in the space around them; so it appears that they have accurate internal models of 3-D space - almost as precise as can be built from their sense data.
2. An object tracking computation can build a precise 3-D model of local space, almost as good as the best Bayesian model that can be built from sense data. Tracking is a good candidate for how animals do 3-D spatial cognition.
3. In spite of the great importance of spatial cognition for fitness and survival, there are no working neural

models of 3-D spatial cognition. This is a serious challenge for neuroscience – especially in insect brains, with fewer than 1 million neurons.

4. Neural spatial memory is too slow and imprecise to support the tracking computation, or other models of spatial cognition.
5. Wave storage of spatial information can give high precision, fast response times and low spatial distortion. A wave is precise enough to support the tracking model.
6. No such wave has been observed in the brain, but it has not been looked for; if it exists, it may have very low amplitude which is hard to detect.
7. The central body of the insect brain has a remarkably preserved shape across all insect species, and its round shape is well suited to hold a 3-D wave excitation.
8. The insect central body has all the connections required for spatial cognition and multi-sensory integration. Significantly, it has few links to olfaction, which is not useful for fast spatial computation.
9. The size of the central body is constant within a factor 2.5 over a range of 40 in insect brain volumes, consistent with the hypothesis that it holds a wave representing space, to the resolution of insect vision.
10. The mammalian thalamus has a remarkably preserved shape across all species, and its round shape is well suited to hold a 3-D wave excitation.
11. Without a wave excitation, the neuroanatomy of the thalamus does not make sense. The same neural computation would be possible if thalamic nuclei migrated out towards cortex, saving net axon length and saving energy consumption.
12. The mammalian thalamus has all the connections required for spatial cognition and multi-sensory integration. Significantly, it has few links to olfaction.
13. The shell-like form of the mammalian TRN only makes sense if the thalamus holds a wave.
14. Spatial steering and binding is an essential cognitive function, but there are no neural models of it with acceptable scaling. A wave model can do signal steering with good scaling.
15. The insect central body and the mammalian thalamus are both ideally placed for sensory signal steering, as the hubs of nearly all incoming sense data.
16. Complex synapses in the insect central body and the thalamus are consistent with a wave-based signal steering role.

These three groups of evidence – concerning memory performance, the insect central body, and the mammalian thalamus – imply that the wave hypothesis merits further investigation.

The list brings out the many parallels between the insect central body and the mammalian thalamus.

## 11. Conclusion: Twin Tracks for Research

Spatial cognition is the primary cognitive function, required before any other function of a brain. Animals move, and they need spatial cognition to control their movements, at every moment of the day. There has been extreme and sustained selection pressure over 500 million years to do it well, and animals do it very well.

Given the primacy of spatial cognition, it is remarkable that there are essentially no working neural computational models of it. I have suggested a reason for this – that neural spatial memory has neither the precision nor the speed to support spatial cognition, as well as we know animals do it. There is an alternative, that spatial memory is held in a wave excitation, in the insect central body or the mammalian thalamus.

There are then two alternative hypotheses; either that spatial short-term memory is held in neural firing rates, in some way we have not yet thought of, or that spatial memory is held in a wave. This suggests a twin-track research agenda, to explore the two hypotheses in parallel.

I have outlined the difficulties facing a neural spatial memory model; others may devise solutions better than I can. This is a worthwhile research agenda, whether or not it succeeds. If it does not succeed, it will at least clarify the nature of the roadblocks, showing the way to better models of spatial memory.

Both research tracks can benefit from experimental work to measure the quality of animals' internal 3-D models of local space – especially in small animals such as insects. I suggest that motion detection is a suitable task to measure the quality of the brain's 3-D model of space, for instance in bees.

The wave hypothesis of spatial cognition raises many questions of biophysics. The core question is: what is the physical nature of the wave? A related question is - why has a wave in the brain not been detected already?

A wave may not yet have been detected because we have not known how to look for it. There is also another reason. The wave would be expected to have evolved in the direction of smaller intensity, to reduce its energy consumption. We know that neurons can detect physical excitations at very low intensities – down to the level of one photon, in the case of light. So we expect a wave in the brain to have evolved to have extremely low intensity. This would makes it very hard to detect – much like the neutrino, which was only detected many years after it was predicted [Pauli 1930, Cowan, Reines et al 1956]; or like dark matter, which is still not identified.

Direct attempts to detect the wave may not be the first priority, until we know more about its nature. There are other possible lines of investigation:

- To look for genes expressed specifically in those parts of the brain, such as the thalamus or the insect central body, which may hold the wave
- If there are genetic correlates of the wave, explore the physical properties of the proteins they encode
- Study the ultrastructure of those parts of the brain, comparatively across more species, to analyse the differences from other brain parts which are not candidates to hold a wave.
- Make computations of brain energy budgets, like the 'exploding thalamus' analysis of section 10, using connectome data to make them more precise, to check whether a wave is necessary to account for the shapes of these brain parts.
- Investigations can be widened to other phyla, such as arachnids, birds and reptiles.
- Studies of the insect brain may be particularly useful because of its greater need for compactness, simplicity and efficiency.
- Build computational models of spatial cognition a Marr's [1982] level of neural implementation, extended by a wave – for instance, in the neural process theory of the Free Energy Principle [Friston 2010]

Many different specialties can contribute to this effort. Searching for a wave in the brain is an exciting green-fields area of research, away from the well-trodden paths of mainstream neuroscience; an area where new ideas can be proposed and discoveries made. If the wave hypothesis were confirmed, it would be a paradigm shift in neuroscience, leading to many new fields of research. That chance makes it worth the attempt.

## References

Chittka, L. (2022) The Mind of a Bee, Princeton University Press, Princeton, NJ

Cowan C.L ; F. Reines; F. B. Harrison; H. W. Kruse; A. D. McGuire (1956). "Detection of the Free Neutrino: a Confirmation". Science. 124 (3212): 103–4.

Denker, J., Schwarz, D., Wittner, B., Solla, S., Howard, R., Jackel, L., et al. (1987). Large automatic learning, rule extraction and generalization. *Complex Syst.* 1, 877–922.

Erman, L. D., Hayes-Roth, F., Lesser, V. R., and Reddy, R. (1980). The HEARSAY-II speech understanding system. Comput. Surv. 12, 213–253.

Feldman, J. (2013). The neural binding problem(s). *Cogn. Neurodyn.* 7, 1–11. doi: 10.1007/s11571-012-9219-8

Friston, K. (2003). Learning and inference in the brain. *Neural Netw.* 16, 1325–1352. doi: 10.1016/j.neunet.2003.06.005

Friston K. (2010) The free-energy principle: a unified brain theory? Nature Reviews Neuroscience

Hailey A. C. and Krubitzer L. (2019) Not all cortical expansions are the same: the coevolution of the neocortex and the dorsal thalamus in mammals, current opinions in neurobiology 56:78

Halassa, M. M., and Sherman, S. M. (2019). Thalamocortical circuit motifs: a general framework. *Neuron* 103, 762–775. doi: 10.1016/j.neuron.2019.06.005

Hebb, D.O. (1949). The Organization of Behavior. New York: Wiley & Sons

Heinze S. et al (2023), the Insect Brain Database, https://insectbraindb.org . Curators of the database are: Berg B.G., Bucher G., el Jundi B, Farnworth M., Gruithuis J., Hartenstein V., Heinze S., Hensgen R., Homberg U., Pfeiffer K., Pfuhl G., Rossler W., Rybak J., Younger M.

Jones, E. G. (2007). The Thalamus, 2nd edition. New York, NY: Cambridge University Press.

Knill, D. C., and Pouget, A. (2004). The Bayesian brain: the role of uncertainty in neural coding and computation. *Trends Neurosci.* 27, 712–719. doi: 10.1016/j.tins.2004.10.007

Lee, T. S., and Mumford, D. (2003). Hierarchical Bayesian inference in the visual cortex. *J. Opt. Soc. Am. A Opt. Image Sci. Vis.* 2, 1434–1448. doi: 10.1364/josaa.20.001434

Llinas, J., and Anthony, R. T. (1993). Blackboard concepts for data fusion applications. *Int. J. Pattern Recognit. Artif. Intell.* 7, 285–308. doi: 10.1142/S0218001493000157

McFadden, J. (2002) Synchronous firing and its influence on the brain's magnetic field. Journal of Consciousness Studies, 9, 23-50.

McCulloch W.S. & W. Pitts (1943), A logical calculus of ideas immanent in nervous activity, Bulletin of Mathematical Biophysics 5, 115

Mumford, D. (1991). On the computational architecture of the neocortex I: the role of the thalamo-cortical loop. Biol. Cybern. 65, 135–145. doi: 10.1007/BF00202389

Murray, S. O., Olshausen, B. A., and Woods, D. L. (2003). Processing shape, motion and three-dimensional shape-from-motion in the human cortex. *Cereb. Cortex* 13, 508–516. doi: 10.1093/cercor/13.5.508

Nii, P. (1986). The blackboard model of problem solving and the evolution of blackboard architectures. *AI Mag.* 7:38. doi: 10.1609/aimag.v7i2.537

Olshausen, B. A., Anderson, C. H., and Van Essen, D. C. (1993). A neurobiological model of visual attention and invariant pattern recognition based on dynamic routing of information. *J. Neurosci.* 13, 4700–4719. doi: 10.1523/JNEUROSCI.13-11-04700.1993

Olshausen, B. A., Anderson, C. H., and Van Essen, D. C. (1995). A multiscale dynamic routing circuit for forming size- and position-invariant object representations. *J. Comput. Neurosci.* 2, 45–62 doi: 10.1007/BF00962707

Pauli, W (1930) "Liebe Radioaktive Damen und Herren" [Dear Radioactive Ladies and Gentlemen].

Rudrauf, D., Bennequin, D., Granic, I., Landini, G., Friston, K., and Williford, K. (2017). A mathematical model of embodied consciousness. J. Theor. Biol. 428, 106–131

Rudrauf D, Sergeant-Perthuis G. Belli O. and Di Marzo Serugendo G. (2022) Modeling the subjective perspective of consciousness and its role in the control of behaviours, Journal of Theoretical Biology 534

Sherman, S. M. (2007). The thalamus is more than just a relay. *Curr. Opin. Neurobiol.* 17, 412–422. doi: 10.1016/j.conb.2007.07.003

Sherman, S. M. and Guillery, R. W (2006) Exploring the Thalamus and its role in Cortical Function, MIT Press, Cambridge, Mass

Spacek J and Lieberman A R (1974) Ultrastructure and three-dimensional organization of synaptic glomeruli in rat somatosensory thalamus. J Anat. Jul; 117(Pt 3): 487–516.

Treisman, A. (1998). Feature binding, attention and object perception. *Philos. Trans. R. Soc. Lond. B Biol. Sci.* 353, 1295–1306. doi: 10.1098/rstb.1998.0284

Treisman, A., and Gelade, G. (1980). A feature integration theory of attention. *Cogn. Psychol.* 12, 97–136. doi: 10.1016/0010-0285(80)90005-5

Worden, R. P. (1995). An optimal yardstick for cognition. *Psycoloquy* 7:1.

Worden, R.P. (2010) Why does the Thalamus stay together?, unpublished paper on ResearchGate

Worden, R.P. (2014) the Thalamic Reticular Nucleus: an Anomaly; unpublished paper on ResearchGate

Worden, R. P. (2020a) Is there a wave excitation in the thalamus?  arXiv:2006.03420

Worden, R. P. (2020b). An Aggregator model of spatial cognition. arXiv 2011.05853.

Worden R.P, Bennett M and Neascu V (2021) The Thalamus as a Blackboard for Perception and Planning, Front. Behav. Neurosci., 01 March 2021, Sec. Motivation and Reward, https://doi.org/10.3389/fnbeh.2021.633872

Worden R.P. (2024a) The Requirement for Cognition in an Equation, http://arxiv.org/abs/2405.08601

Worden R.P. (2024b) Three-dimensional Spatial Cognition: Bees and Bats, http://arxiv.org/abs/2405.09413